\documentclass[11pt]{article}

\usepackage{palatino}
\usepackage{mathpazo}
\usepackage{graphicx}
\usepackage[margin=1in]{geometry}
\usepackage{amssymb, amsthm, amsmath}

\newtheorem{theorem}{Theorem}
\newtheorem{lemma}[theorem]{Lemma}

\def\<{\langle}
\def\>{\rangle}
\def\be{\begin{equation}}
\def\ee{\end{equation}}

\newcommand{\comment}[1]{}

\newcommand{\h}{\mathcal{H}}

\newcommand{\ket}[1]{\ensuremath{\left|#1\right\rangle}}
\newcommand{\op}[2]{|#1\rangle \langle #2|}
\newcommand{\Tr}{\mathop{\mathrm{tr}}\nolimits}

\title{\bf\LARGE Maximum privacy without coherence, zero-error}
\author{
  Debbie Leung\footnote{%
    Institute for Quantum Computing and
    Department of Combinatorics and Optimization,
    University of Waterloo,
    Waterloo, Ontario, Canada.}
  \and
  Nengkun Yu\footnote{
Institute for Quantum Computing and
    Department of Combinatorics and Optimization,
    University of Waterloo,
    Waterloo, and
Department of Mathematics $\&$ Statistics, University of Guelph, Guelph, Ontario, Canada
  }
}
\date{03 September, 2015}  

\begin{document}
\maketitle

\begin{abstract}
We study the possible difference between the quantum and the private
capacities of a quantum channel in the zero-error setting.  For a
family of channels introduced by \cite{Leung-Li-Smith-Smolin-14}, we
demonstrate an extreme difference: the zero-error quantum capacity is
zero, whereas the zero-error private capacity is maximum given the
quantum output dimension.

\end{abstract}

\section{Introduction}
\label{intro}

Given Alice, a sender, and Bob, a receiver, any communication from
Alice to Bob can be modeled by a quantum channel $\mathcal{N}$.
Various capacities of the quantum channel $\mathcal{N}$ can be defined
to quantify its capability for communicating different types of data.
The quantum capacity $Q(\mathcal{N})$, measured in qubits per channel
use, establishes the maximum rate for transmitting quantum
information and how well we can perform quantum error correction.
The private capacity $\mathcal{P}(\mathcal{N})$, in bits per channel use,
gives the maximum rate of {\em private} classical communication.
Errors that become negligible as the number of channel uses increases
are allowed in the above definitions.

Understanding the relation between the quantum and the private
capacities is an essential task in quantum Shannon theory.  In
\cite{HHHO03}, some channels $\mathcal{N}$ are found for which
$Q(\mathcal{N}) = 0$ but $P(\mathcal{N}) > 0$, breaking a long-held
intuition that coherence is necessary for privacy.  In
\cite{Leung-Li-Smith-Smolin-14}, a class of channels with
$Q(\mathcal{N})\leq 1$ and $P(\mathcal{N})=\log d$ is presented, where
$d^2$ is the input dimension and log is taken base 2.  As $d$
increases, these channels saturate an upper bound for
$P(\mathcal{N})-Q(\mathcal{N})$ thus approximately realizing the
largest possible separation between the two capacities.

Introduced by Shannon in 1956 \cite{Shannon56}, the zero-error
capacity characterizes the optimal achievable communication rate of a
noisy channel when information must be transmitted without any
error. The zero-error capacity has deep connections to combinatorial
optimization and it plays an essential role in graph theory and
communication complexity theory \cite{Lovasz79,Alon98}.  Quite
recently, the notion of zero-error capacity has been introduced for
quantum channels \cite{MA05,MACA06,Beigi-Shor07}, and many further
interesting results are found
\cite{Duan09,CubittChenHarrow11,CubittSmith12,DuanSeveriniWinter13,Shirokov-Shulman-14,Shirokov-Shulman-15,Shirokov-15}.
In particular, some channels cannot transmit quantum data perfectly
given only one use, but have a large transmission rate with two uses.
Also, there exist channels with no zero-error capacity but whose joint
zero-error capacity is positive, a phenomenon called superactivation.
In fact, quantum channels with no zero-error {\em classical}
capacities are found to have joint zero-error {\em quantum} capacities
(that activates all of zero-error classical, private, and quantum
capacities)!  Superactivation is impossible for classical channels.
Various assisted communication scenarios
have also been studied but they are out of the current scope.

We denote the zero-error quantum and private capacities for a quantum
channel $\mathcal{N}$ as $Q_0(\mathcal{N})$ and $P_0(\mathcal{N})$
respectively.  Zero-error private classical communication requires
perfect data transmission such that no one but the receiver
gains any information on the data.  Clearly $Q_0(\mathcal{N})\leq
Q(\mathcal{N})\leq \mathcal{P}(\mathcal{N})$ and $Q_0(\mathcal{N})\leq
P_0(\mathcal{N})\leq \mathcal{P}(\mathcal{N})$.

In this paper, we study the zero-error quantum capacity of the channels
introduced in \cite{Leung-Li-Smith-Smolin-14}, and demonstrate an
exact extreme separation.  For these channels, $P_0(\mathcal{N}) =
\log d$ and $Q_0(\mathcal{N})=0$.  In other words, each of these
channels has no capacity to transmit quantum information perfectly,
even it has full ability to distribute private information perfectly.


\section{Preliminaries}
\label{primer}

In this section, we discuss some background and notation in quantum
information and zero error capacities for a quantum channel.  Readers
familiar with these subjects can proceed to the next section.

A {\it complex Euclidean space} refers to any finite dimensional inner
product space over the complex numbers.  Let $\h_A$ and $\h_B$ be
arbitrary complex Euclidean spaces. A pure quantum state of $\h_A$ is
a normalized vector $\ket{\psi}\in \h_A$.

The space of linear operators mapping $\h_A$ to $\h_B$ is denoted by
$\mathcal{L}(\h_A,\h_B)$, while $\mathcal{L}(\h_A)$ is the shorthand
for $\mathcal{L}(\h_A,\h_A)$.  $I_{\h_A}$ is used to denote the
identity operator on $\h_A$.  The adjoint of $M\in\mathcal{L}(\h_A)$
is denoted by $M^{\dag}$.  The notation $M\geq 0$ means that $M$ is
hermitian, $M=M^{\dag}$, and is positive semidefinite.

With respect to a fixed basis, the complex conjugate of a state
$\ket{\alpha}$ and a linear operator $M$ are denoted as
$\ket{\alpha^c}$ and $M^c$, respectively.

A general quantum state of $\h_A$ is characterized by its density operator
$\rho\in \mathcal{L}(\h_A)$, which is a positive semidefinite
operator with trace one on $\h_A$.  We denote the set of density
matrices as $\mathcal{D}(\h_A)$.
The density operator of a pure
state $\ket{\psi}$ is simply the projector $\psi:=\op{\psi}{\psi}$.
The support of $\rho$, denoted by $supp(\rho)$, is the vector space
spanned by the eigenvectors of $\rho$ with positive eigenvalues.  More
concretely, if $\rho$ has spectral decomposition
$\rho=\sum_{k=1}^n p_k\op{\psi_k}{\psi_k}$, where $0<p_k\leq 1$ and
$\sum_{k=1}^n p_k=1$. Then $supp(\rho)=span\{\ket{\psi_k}:1\leq k\leq
n\}$.  The null space of any $M\geq 0$ is the orthogonal complement of
$supp(M)$.

A nonzero positive semidefinite operator
$E \in \mathcal{L}(\h_A\otimes \h_B)$
is said to be a PPT operator (or simply PPT) if $E^{\Gamma_{\h_A}}\geq0$,
where ${\Gamma_{\h_A}}$ denotes the partial transpose with respect
to the system $\h_A$, i.e.,
\begin{equation}
(\op{ij}{kl})^{\Gamma_{\h_A}}=\op{kj}{il}.
 \end{equation}
While ${\Gamma_{\h_A}}$ is basis dependent, the property being PPT
is not.
For simplicity, we will specify the system $\h_A$ for $\Gamma_{\h_A}$
in the text and denote the operation with the shorthand $\Gamma$.

A quantum channel $\mathcal{N}$ from Alice to Bob is a completely
positive trace preserving linear map from the input state space of
Alice $\mathcal{D}(\h_A)$ to the output state space of Bob
$\mathcal{D}(\h_B)$.  There are several characterizations of quantum
channels (see \cite{NC00} chapter 8 or \cite{Watrous-notes}).  We use
the isometric extension for a channel, $\mathcal{N}(\rho) = \Tr_{\h_E}
U \rho U^\dagger$, where $U \in \mathcal{L}(\h_A, \h_B \otimes \h_E)$
is an isometry mapping the input to the output space for Bob and some
auxilary space called the ``environment'', and $\Tr_{\h_E}$ denote the
partial trace over $\h_E$.  The isometry $U$ is unique up to left
multiplication by another isometry acting on $\h_E$, and this degree
of freedom has no physical effect on our analysis, and can be chosen to
facilitate it.  The complementary channel $\mathcal{N}^c(\rho) = \Tr_{\h_B}
U \rho U^\dagger$ describes what information leaks to the environment.


The notion of zero-error quantum capacity can be introduced as
follows. Let $\alpha^q(\mathcal{N})$ be the maximum integer $k$ such
that there is a $k$-dimensional subspace $\h_A'$ of $\h_A$ that can be
perfectly transmitted through $\mathcal{N}$. That is, there is a
recovery quantum channel $\mathcal{R}$ from
$\mathcal{D}(\h_B)$ to $\mathcal{D}(\h_{A'})$ so that
$(\mathcal{R}\circ \mathcal{N})(\psi)=\psi$ for any
$\ket{\psi}\in \h_{A'}$ (recall $\psi = \op{\psi}{\psi}$).
Then, $\log_2 \alpha^q(\mathcal{N})$ represents the maximum number of
qubits one can send perfectly by one use of $\mathcal{N}$. The
{\em zero-error quantum capacity} of $\mathcal{N}$, $Q_0(\mathcal{N})$,
is defined as:
\begin{equation}\label{q0}
Q_0(\mathcal{N})=\sup_{n\geq 1}\frac{\log_2 \alpha^q(\mathcal{N}^{\otimes n})}{n}.
\end{equation}

The main difficulty of evaluating the zero-error capacity of a quantum
channel is that there is no upper bound on the required number of uses
$n$ in evaluating the above expression.  This remains the case even
for the simpler problem of determining whether $Q_0(\mathcal{N}) = 0$.
For example, in \cite{Shirokov-Shulman-14}, for any integer $k$, the
authors found a channel $\mathcal{N}$ for which
$\alpha^q(\mathcal{N}) = 1$ but
$\alpha^q(\mathcal{N}^{\otimes 2}) \geq k$.
They also found a channel $\mathcal{N}$ for which the $k$-shot capacity
vanishes but $Q_0(\mathcal{N}) > 0$.
Furthermore, superactivation is possible (see section \ref{intro})
and \cite{CubittSmith12} exhibits an extreme example
in which channels $\mathcal{N}_1$, $\mathcal{N}_2$
have no zero-error classical capacity but
$Q_0(\mathcal{N}_1 \otimes \mathcal{N}_2) >0$.
Fortunately, for our
purpose, we can invoke the following lemma from \cite{CubittSmith12}.
\begin{lemma}\label{lem:quantum_zero-error}
  Let $\mathcal{N}:\mathcal{D}(\h_A) \rightarrow \mathcal{D}(\h_B)$
  be a quantum channel.  One can
  transmit quantum information without error through a single use of
  $\mathcal{N}$ if and only if there are states $\ket{\alpha}$ and
  $\ket{\beta}$ such that
  \begin{equation}
    \Tr\left[ \mathcal{N}(\op{\alpha}{\alpha})\,\mathcal{E}(\op{\beta}{\beta}) \right] = 0
  \end{equation}
  and
  \begin{equation}
    \Tr\left[ \mathcal{N}(\op{\alpha+\beta}{\alpha+\beta})\,
              \mathcal{N}(\op{\alpha-\beta}{\alpha-\beta}) \right] = 0.
  \end{equation}
\end{lemma}
where $\ket{{\alpha\pm\beta}} = 1/\sqrt{2}(\ket{\alpha} \pm \ket{\beta})$.

Private communication via a memoryless classical channel and quantum key
distribution are well established subjects.  Private classical
communication of a quantum channel has more recently been formally
introduced in \cite{Devetak05}.  The private capacity of $\mathcal{N}$
measures the maximum rate of reliable classical data transmission via
$\mathcal{N}$ while keeping the output of the complementary channel
independent of the data.  In \cite{Devetak05}, an expression for
the private capacity is derived,
\be
\mathcal{P}(\mathcal{N}) = \max_{\mathcal{E}} \frac{1}{n}  \left[
I(X:B_1 \cdots B_n) -  I(X:E_1 \cdots E_n) \right]
\label{eq:pcap}
\ee
where $B_i, E_i$ are the output and environment spaces for the
$i^{\rm th}$ channel use, $I(C:D) = S(C) + S(D) - S(CD)$ is
the quantum mutual information between $C$ and $D$ evaluated
on the state of $CD$, $S(\cdot)$ denotes the von Neumann entropy,
and $\mathcal{E} = \{p_x, \rho_x\}$ is a general ensemble of
possibly mixed states $\rho_x$ on the $n$ input spaces $A_1 \cdots A_n$.
The expressions
$I(X:B_1 \cdots B_n)$,
$I(X:E_1 \cdots E_n)$
are evaluated on
$\sum_{x} p_x |x\>\<x|_X \otimes \mathcal{N}^{\otimes n}(\rho_x)$
and
$\sum_{x} p_x |x\>\<x|_X \otimes \mathcal{N}^{c \otimes n}(\rho_x)$
respectively.
Once again, the requirement to optimize over $n$ in the capacity
expression Eq.~(\ref{eq:pcap}) is an obstacle for evaluating the
private capacity in general.  However, useful lower bounds and
properties of the private capacity can still be inferred from
Eq.~(\ref{eq:pcap}).

Consider any quantum channel $\mathcal{N}$ with a quantum output $B$
and a classical output $C$.  We use Eq.~(\ref{eq:pcap}) to show
that $\mathcal{P}(\mathcal{N}) \leq \log_2 \dim(B)$.  We first
consider the one shot case.
\be
I(X:BC) - I(X:E) = I(X:BC) - I(X:EC) =
\sum_c p_c [I(X:B|C=c) - I(X:E|C=c)] \,.
\ee
The first equality comes from the classicality of $C$, that the
environment $E$ already has a copy of the information.  The second
equality follows from decomposing the von Neuman entropy of a
quantum-classical system \cite{NC00}.  If we optimize the expression in
the square brackets, we obtain the one-shot private capacity
for $\mathcal{N}$ conditioned on $C=c$, which is upper bounded by
$\log_2 \dim(B)$.  The argument for the $n$-shot case is identical,
with $B,C,E$ replaced by $n$-tuples.

\section{Zero-error Quantum Capacity of $\mathcal{N}_d$}
\label{nd}

In this section, we first describe the family of channels that will
exhibit the extreme separation between the zero-error quantum and
private capacities.  Then, we derive those capacities.

The family of channels $\mathcal{N}_d$ introduced in
\cite{Leung-Li-Smith-Smolin-14} can be schematically
summarized as follows:

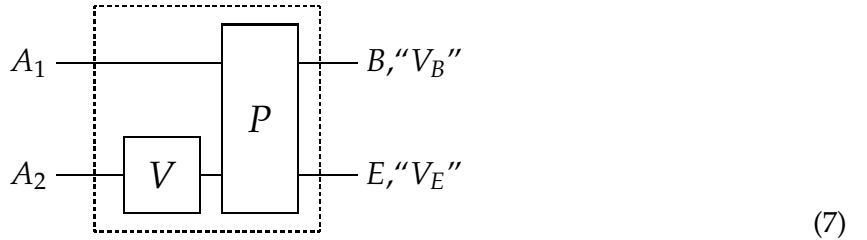
\begin{equation}
\setlength{\unitlength}{0.5mm}
\centering
\begin{picture}(130,58)
\put(30,0){\dashbox(60,60){}}
\put(05,10){\makebox(15,10){\large{$A_2$}}}
\put(05,40){\makebox(15,10){\large{$A_1$}}}
\put(20,15){\line(1,0){18}}
\put(84,45){\line(1,0){16}}
\put(20,45){\line(1,0){44}}
\put(38,05){\framebox(20,20){\Large{$V$}}}
\put(58,15){\line(1,0){6}}
\put(64,05){\framebox(20,50){\Large{$P$}}}
\put(84,15){\line(1,0){16}}
\put(100,10){\makebox(30,10){\large{$E$,``$V_E$''}}}
\put(100,40){\makebox(30,10){\large{$B$,``$V_B$''}}}
\end{picture}
\label{eq:Channel}
\end{equation}

For each integer $d \geq 2$, we define the channel $\mathcal{N}_d$
which has two input registers $A_1$ and $A_2$, each of dimension $d$.
A unitary operation $V$ is applied to $A_2$, followed by a controlled
phase gate $P = \sum_{i,j} \omega^{ij}\op{i}{i}\otimes \op{j}{j}$
acting on $A_1 A_2$, where $\omega$ is a primitive $d^{\rm th}$ root of
unity.  Bob receives only $A_1$ (now relabeled as $B$) and ``$V_B$'', which
denotes a classical register with a description of $V$.  The $A_2$
register is discarded.  The complementary channel has outputs $A_2$
(relabeled as $E$) and ``$V_E$'' which also contains a description of $V$.
The isometric extension is given by
\[
U_d \ket{\psi}_{A_1A_2} = \sum_V \! \sqrt{{\rm pr}(V)} \,
\left( P \,(I\otimes V) \ket{\psi}_{A_1A_2} \right)
\otimes \ket{V}_{V_B}\otimes \ket{V}_{V_E} \,.
\]
Here, $V$ is drawn from any exact unitary 2-design
$\mathcal{G}=\{g_1,g_2,\cdots,g_m\}$ (such as the Clifford group,
see \cite{CLLW15} and the references therein).


It was shown in \cite{Leung-Li-Smith-Smolin-14} that $P(\mathcal{N}_d)
= \log d$.  The method given by \cite{Leung-Li-Smith-Smolin-14} to
transmit private classical data has no error and has perfect secrecy
so $P_0(\mathcal{N}_d) = \log d$.  To be self-contained, we provide
a quick argument here.  Suppose the input into $A_2$ is half of a
maximally entangled state
$\ket{\Phi}=\frac{1}{\sqrt{d}}\sum_{i}\ket{i}_{A_2}\ket{i}_{A_3}$ where
$A_3$ stays in Alice's possession.  By
the transpose trick, the unitary operations $V$ and $P$ can be
replaced by unitary operations acting on $A_1$ and $A_3$ without
changing the final state on $B,E,A_3,V_B,V_E$.  So, the output of the
complementary channel ($E, V_E$) is independent of the input.
Moreover, $\mathcal{N}_d(\op{i}{i} \otimes I/d) = \op{i}{i}$.  So
$\log d$ bits can be transmitted perfectly and secretly.

Furthermore, \cite{Leung-Li-Smith-Smolin-14} also shows that
$Q(\mathcal{N}_d) \leq 1$.  Intuitively, superposition of states in
system $A_1$ will be heavily decohered by the $P$ gate, because error
correction is ineffective due to the random unitary $V$.  However,
\cite{Leung-Li-Smith-Smolin-14} finds that $Q(\mathcal{N}_d) \geq 0.61$
for large $d$.

This motivates the current study, to demonstrate an extreme separation
of $P_0$ and $Q_0$ using the channels $\mathcal{N}_d$.  Our main
result is that, no finite number of uses of $\mathcal{N}_d$ can be
used to transmit one qubit with zero error.  This implies in
particular $Q_0(\mathcal{N}_d) = 0$, while $P_0(\mathcal{N}_d) = \log
d$, attaining the extremes allowed by the quantum output dimension
(see end of section \ref{primer}.)

In \cite{Duan09,DuanSeveriniWinter13,Shirokov-Shulman-14}, the
non-commutative graph of a quantum channel is defined and used to
study several different zero-error capacities.  In particular,
\cite{Shirokov-Shulman-14} derived sufficient conditions for the
impossibility of superactivation of zero-error quantum capacity of a
channel in terms of properties of the single-copy non-commutative
graph.  If a channel satisfies any of these conditions, and has no
one-shot zero error quantum capacity, then it has no zero-error
quantum capacity by induction.  However, we show in the appendix that
the non-commutative graph of $\mathcal{N}_d$ does not satisfy any of
these conditions, along with a discussion of its non-commutative
graph.

Our main technical result is a characterization of pairs of input
states whose orthogonality is preserved by $n$ uses of the channel.
\begin{theorem}\upshape \label{th1}
Let $n$ be any positive integer,
$\ket{\psi_1}=\sum_{i_1,\cdots,i_n}\ket{i_1,\cdots,i_n}\ket{\alpha_{i_1,\cdots,i_n}}$,
and
$\ket{\psi_2}=\sum_{i_1,\cdots,i_n}$ $\ket{i_1,\cdots,i_n}\ket{\beta_{i_1,\cdots,i_n}}$
be two arbitrary pure state inputs for $\mathcal{N}_d^{\otimes n}$.
Then,
$\Tr[\mathcal{N}_d^{\otimes n}(\psi_1)\mathcal{N}_d^{\otimes n}(\psi_2)]=0$
if and only if
at most one of $\ket{\alpha_{i_1,\cdots,i_n}}$ and
$\ket{\beta_{i_1,\cdots,i_n}}$ is nonzero for each tuple $(i_1,\cdots,i_n)$.
\end{theorem}
In other words, states suitable for transmitting classical information
through $\mathcal{N}_d^{\otimes n}$ without any error have no ``overlap" in the
computational basis of $A_1^{\otimes n}$.

We first state the consequence of Theorem \ref{th1} and then we
will return to prove it.

\begin{theorem}\upshape \label{main}
For any positive integer $n$, $\mathcal{N}_d^{\otimes n}$ cannot
transmit a qubit with zero error.  In particular, this implies
$Q_0(\mathcal{N}_d)=0.$
\end{theorem}
{\bf Proof (theorem \ref{main}).} Suppose by contradiction, some
$n$ uses of $\mathcal{N}_d$
can be used to transmit a $2$-dimensional subspace spanned by a basis
$\{\ket{\psi_1},\ket{\psi_2}\}$, where
$\ket{\psi_1}=\sum_{i_1,\cdots,i_n}\ket{i_1,\cdots,i_n}\ket{\alpha_{i_1,\cdots,i_n}}$,
and
$\ket{\psi_2}=\sum_{i_1,\cdots,i_n}$ $\ket{i_1,\cdots,i_n}\ket{\beta_{i_1,\cdots,i_n}}$.
According to Lemma \ref{lem:quantum_zero-error},
\begin{equation}
    \Tr\left[ \mathcal{N}_d^{\otimes n}(\psi_1)\,
              \mathcal{N}_d^{\otimes n}(\psi_2) \right] = 0
  \end{equation}
  and
  \begin{equation}
    \Tr\left[
    \mathcal{N}_d^{\otimes n}(\op{\psi_1{+}\psi_2}{\psi_1{+}\psi_2})\,
    \mathcal{N}_d^{\otimes n}(\op{\psi_1{-}\psi_2}{\psi_1{-}\psi_2})
    \right] = 0.
  \end{equation}
Invoking Theorem \ref{th1} on the conditions above, for each
$i_1,\cdots,i_n$, at most one of $\ket{\alpha_{i_1,\cdots,i_n}}=0$ and
$\ket{\beta_{i_1,\cdots,i_n}}$ is nonzero, and at most one of
$(\ket{\alpha_{i_1\cdots i_n}}+\ket{\beta_{i_1,\cdots,i_n}})$ and
$(\ket{\alpha_{i_1\cdots i_n}}-\ket{\beta_{i_1,\cdots,i_n}})$ is
nonzero, which implies
$\ket{\alpha_{i_1\cdots i_n}} = \ket{\beta_{i_1,\cdots,i_n}} = 0$.
Then, $\ket{\psi_1}=\ket{\psi_2}=0$ a contradiction.
\hfill $\blacksquare$


We now turn to a proof for Theorem \ref{th1}.  We first consider the
simpler one-shot case to illustrate the main ideas without the burden
of the $n$-shot notations.  Then, we prove Theorem \ref{th1} with
similar techniques.

\begin{lemma}\upshape \label{lem1}
Let $\ket{\psi_1}=\sum_{i}\ket{i}\ket{\alpha_{i}}$ and
$\ket{\psi_2}=\sum_{i}\ket{i}\ket{\beta_{i}}$ be two possible
pure input states for $\mathcal{N}_d$.  Then,
$\Tr[\mathcal{N}_d(\psi_1)\mathcal{N}_d(\psi_2)]=0$
   if and only if at most
  one of $\ket{\alpha_{i}}$ and $\ket{\beta_{i}}$ is nonzero for each
  $i$.
\end{lemma}
{\bf Proof (lemma \ref{lem1}).}
We first rephrase the condition $\Tr[\mathcal{N}_d(\psi_1)\mathcal{N}_d(\psi_2)]=0$.

Recall that $V$ is chosen from an exact unitary $2$-design
$\mathcal{G}=\{g_1,g_2,\cdots,g_m\}$.
We rewrite the gate $P = \sum_{i} \op{i}{i} \otimes Z_i$ where
$Z=\sum_{k}\omega^{k}\op{k}{k}$, and $Z_l=Z^l$.  So, for $V=g_j$,
\begin{eqnarray*}
&&P(I\otimes V)\ket{\psi_1}= \sum_{i}\ket{i} \left(Z_ig_j\ket{\alpha_{i}}\right),\\
&&P(I\otimes V)\ket{\psi_2}= \sum_{i}\ket{i} \left(Z_ig_j\ket{\beta_{i}}\right),
\end{eqnarray*}
where the first and second systems are $B$ and $E$ respectively.  Then
reducing the above states to $B$ gives respectively
\begin{eqnarray*}
&&\rho^{(j)}=\sum_{i,k} \, \rho_{i,k}^{(j)} \, \op{k}{i} ~~~{\rm with}~~~
 \rho_{i,k}^{(j)}=\langle \alpha_i|g_j^{\dag}Z_{k-i}g_j|\alpha_k\rangle,\\
&&\sigma^{(j)}=\sum_{i,k} \, \sigma_{i,k}^{(j)} \, \op{k}{i}~~~{\rm with}~~~
 \sigma_{i,k}^{(j)}=\langle \beta_i|g_j^{\dag}Z_{k-i}g_j|\beta_k\rangle.
\end{eqnarray*}
Their trace inner product can be rephrased:
\begin{eqnarray*}
\Tr(\rho^{(j)\dag}\sigma^{(j)})&=&\Tr(\rho^{(j)}\sigma^{(j)})
=\sum_{i,k} \rho_{i,k}^{(j)} \sigma_{k,i}^{(j)}
=\sum_{i,k} \rho_{i,k}^{(j)} \sigma_{i,k}^{(j)*}\\
&=&\sum_{i,k} \< \alpha_i|g_j^{\dag}Z_{k-i}g_j|\alpha_k \>
              \< \beta_i^c|g_j^{\dag c}Z_{k-i}^{c}g_j^c|\beta_k^c \> \\
&=&\sum_{i,k} \< \alpha_i| \< \beta_i^c|(g_j^{\dag}Z_{k-i}g_j)
                 \otimes(g_j^{\dag c}Z_{k-i}^{c}g_j^c)|\alpha_k\> |\beta_k^c\> \\
&=& \< x|A^{(j)}|x\> \,,
\end{eqnarray*}
where
\begin{eqnarray*}
A^{(j)}&=&\sum_{i,k}\op{i}{k}\otimes
   (g_j^{\dag}Z_{k-i}g_j)\otimes(g_j^{\dag c}Z_{k-i}^{c}g_j^c)\,, ~{\rm and}\\
\ket{x}&=&\sum_i\ket{i}\ket{\alpha_i}\ket{\beta_i^c}.
\end{eqnarray*}
Let $A:=\mathbb{E}_j A^{(j)}$. By the construction of $\mathcal{N}_d$,
\be
\Tr[\mathcal{N}_d(\psi_1)\mathcal{N}_d(\psi_2)]=0
~\Longleftrightarrow~ \forall j,~\Tr(\rho^{(j)\dag}\sigma^{(j)})=0
~\Longleftrightarrow~ \mathbb{E}_j \Tr(\rho^{(j)\dag}\sigma^{(j)})=0
~\Longleftrightarrow~ \langle x|A|x\rangle=0.
\nonumber
\ee

Having rephrased the condition
$\Tr[\mathcal{N}_d(\psi_1)\mathcal{N}_d(\psi_2)]=0$ as $\<x|A|x\>=0$,
we show below that the latter implies $\ket{x}=0$.  This is done by
first evaluating $A$ which has very simple structure, then analyzing its
null space, and showing that the null space contains no nonzero state of
the form given by $|x\>$.


One can calculate $A$ directly because $\{g_j\}$ is an exact unitary
$2$-design \cite{DCEL09,CLLW15}:
\begin{eqnarray}
A&=&\mathbb{E}_j A^{(j)}\\
&=& \sum_{i,k}\op{i}{k}\otimes \mathbb{E}_j (g_j^{\dag}\otimes g_j^{\dag c})
                                (Z_{k-i}\otimes Z_{i-k})(g_j\otimes g_j^c)\\
&=& \sum_{i,i}\op{i}{i}\otimes I
       + \sum_{i \neq k} \op{i}{k}\otimes \mathbb{E}_j
   (g_j^{\dag}\otimes g_j^{\dag c})(Z_{k-i}\otimes Z_{i-k})(g_j\otimes g_j^c) \,.
\end{eqnarray}
Here, for $i\neq k$, we have
$\mathbb{E}_j (g_j^{\dag}\otimes g_j^{\dag c})
 (Z_{k-i}\otimes Z_{i-k})(g_j\otimes g_j^c)
= -\frac{1}{d^2-1}(I-\Phi) + \Phi$,
where $\ket{\Phi}=\frac{1}{\sqrt{d}}\sum_{i}\ket{i}\ket{i}$ is the
maximally entangled state, $\{\Phi,I-\Phi\}$ form a maximal set of
invariances for the averaging, and the coefficients
$-\frac{1}{d^2-1}$ and $1$ come from evaluating
$\frac{1}{d^2-1}\Tr [Z_{k-i}\otimes Z_{i-k} (I{-}\Phi)]$ and
$\Tr [Z_{k-i}\otimes Z_{i-k} \Phi]$ respectively.
Therefore,
\begin{eqnarray}
A&=& \sum_{i,i}\op{i}{i}\otimes I + \sum_{i \neq k} \op{i}{k}\otimes
     \left(-\frac{1}{d^2-1}(I-\Phi)+\Phi\right)
\\
 &=& \sum_{i,k}\op{i}{k}\otimes (a_{i,k}(I-\Phi)+\Phi),
\end{eqnarray}
where $a_{i,i}= 1$ and $a_{i,k} = -\frac{1}{d^2-1}$ for $i \neq k$.

Having found
an explicit expression for $A$, we analyze the support of $A$ as follows:
\begin{eqnarray*}
A&=&\sum_{i,k}\op{i}{k}\otimes \left( a_{i,k}(I-\Phi)+\Phi \right)\\
&=& \left( \sum_{i,k} a_{i,k} \op{i}{k} \right) \otimes
   (I-\Phi)+ d \; \op{\nu}{\nu} \otimes \Phi \,,
\end{eqnarray*}
where $\ket{\nu}=\frac{1}{\sqrt{d}}\sum_i\ket{i}$.
Notice that $\sum_{i,k}a_{i,k}\op{i}{k}\geq 0$, so the projector onto
the support of $A$ is
$$
I\otimes (I-\Phi)+\op{\nu}{\nu}\otimes \Phi.
$$
So the null space of $A$ is spanned by $\ket{\mu}\otimes \ket{\Phi}$
where $\ket{\mu}$ is any vector orthogonal to $\ket{\nu}$.

We now show that $\ket{x}=0$.
Suppose by contradiction that $\ket{x}\neq 0$.
Since $\<x|A|x\>=0$, we have $\ket{x}=\ket{\mu}\otimes
\ket{\Phi}$ for some $\ket{\mu}\neq0$.
But $\ket{x}=\sum_i\ket{i}\ket{\alpha_i}\ket{\beta_i^c}$.
For each $l \in \{1,\cdots,d\}$,
$$(\<l| \otimes I \otimes I) |x\> = \ket{\alpha_l}\ket{\beta_l^c} = \<l|\mu\> |\Phi\>,$$
which is a contradiction unless $\<l|\mu\>=0$.  But now $|\mu\>=0$ which is
a contradiction.

Putting the above together,
$\Tr[\mathcal{N}_d(\psi_1)\mathcal{N}_d(\psi_2)]=0$ if and only if
$\langle x|A|x\rangle=0$ if and only if
at most one of $\ket{\alpha_i}$ and $\ket{\beta_i^c}$ is nonzero for each $i$.
\hfill $\blacksquare$

We are now going to prove Theorem \ref{th1}, using similar techniques.

{\bf Proof (theorem \ref{th1}).}
Consider two arbitrary pure input states for $n$ uses of $\mathcal{N}_d$,
$\ket{\psi_1}=\sum_{i_1,\cdots,i_n}$ $\ket{i_1\cdots i_n}\ket{\alpha_{i_1\cdots
    i_n}}$ and $\ket{\psi_2}=\sum_{i_1,\cdots,i_n}\ket{i_1\cdots
  i_n}\ket{\beta_{i_1,\cdots,i_n}}$. For
$V_1\otimes\cdots\otimes V_n=g_{j_1}\otimes\cdots\otimes g_{j_n}$, we
have
\begin{eqnarray*}
&&P^{\otimes n }(I^{\otimes n}\otimes V_1\otimes\cdots\otimes V_n)\ket{\psi_1}=\sum_{i_1,\cdots i_n}\ket{i_1\cdots i_n}(Z_{i_1}\otimes \cdots \otimes Z_{i_n})(g_{j_1}\otimes \cdots \otimes g_{j_n})\ket{\alpha_{i_1\cdots i_n}},\\
&&P^{\otimes n }(I^{\otimes n}\otimes V_1\otimes\cdots\otimes V_n)\ket{\psi_2}=\sum_{i_1,\cdots i_n}\ket{i_1\cdots i_n}(Z_{i_1}\otimes \cdots \otimes Z_{i_n})(g_{j_1}\otimes \cdots \otimes g_{j_n})\ket{\beta_{i_1\cdots i_n}}.
\end{eqnarray*}
Then the corresponding output states on $B_1 \cdots B_n$ are
\begin{eqnarray*}
&&\rho^{(j_1,\cdots,j_n)}=
   \sum_{i_1\cdots i_n,k_1\cdots k_n} |k_1\cdots k_n\>\<i_1\cdots i_n|  \;
   \rho_{i_1\cdots i_n,k_1\cdots k_n}^{(j_1,\cdots,j_n)} \\
&&\sigma^{(j_1,\cdots,j_n)}=
   \sum_{i_1\cdots i_n,k_1\cdots k_n} |k_1\cdots k_n\>\<i_1\cdots i_n|  \;
   \sigma_{i_1\cdots i_n,k_1\cdots k_n}^{(j_1,\cdots,j_n)} \,,
\end{eqnarray*}
where
\begin{eqnarray*}
&&\rho_{i_1\cdots i_n,k_1\cdots k_n}^{(j_1,\cdots,j_n)}=\langle \alpha_{i_1\cdots i_n}|(g_{j_1}^{\dag} \otimes \cdots \otimes g_{j_n}^{\dag})(Z_{k_1-i_1}\otimes \cdots \otimes Z_{k_n-i_n})(g_{j_1}\otimes \cdots \otimes g_{j_n})|\alpha_{k_1 \cdots k_n}\rangle,\\
&&\sigma_{i_1\cdots i_n,k_1\cdots k_n}^{(j_1,\cdots,j_n)}=\langle \beta_{i_1\cdots i_n}|(g_{j_1}^{\dag}\otimes \cdots \otimes g_{j_n}^{\dag})(Z_{k_1-i_1}\otimes \cdots \otimes Z_{k_n-i_n})(g_{j_1}\otimes \cdots \otimes g_{j_n})|\beta_{k_1 \cdots k_n}\rangle.
\end{eqnarray*}

As in the one-shot case,
$$\Tr[\mathcal{N}_d^{\otimes n}(\psi_1)\mathcal{N}_d^{\otimes n}(\psi_2)]=0
~\Longleftrightarrow~
\mathbb{E}_{j_1,\cdots,j_n} \Tr(\rho^{(j_1,\cdots,j_n)\dag}\sigma^{(j_1,\cdots,j_n)})=0
~\Longleftrightarrow~
\<x|A^{\otimes n}|x\> = 0\,,$$
where
$\ket{x}=\sum_{i_1,\cdots,i_n}\ket{i_1\cdots i_n}
 \ket{\alpha_{i_1\cdots i_n}}\ket{\beta_{i_1\cdots i_n}^c}$
and $A$ is as defined in the one-shot case.  To verify the last
equivalence:
\begin{eqnarray*}
 &&\mathbb{E}_{j_1,\cdots,j_n} \Tr(\rho^{(j_1,\cdots,j_n)\dag}\sigma^{(j_1,\cdots,j_n)})\\
 &=&\mathbb{E}_{j_1,\cdots,j_n} \sum_{i_1\cdots i_n,k_1\cdots k_n} \rho_{i_1\cdots i_n,k_1\cdots k_n}^{(j_1,\cdots,j_n)}\sigma_{i_1\cdots i_n,k_1\cdots k_n}^{(j_1,\cdots,j_n)*}\\
 &=& \mathbb{E}_{j_1,\cdots,j_n} \sum_{i_1\cdots i_n,k_1\cdots k_n} \langle \alpha_{i_1\cdots i_n}|(g_{j_1}^{\dag} \otimes \cdots \otimes g_{j_n}^{\dag})(Z_{k_1-i_1}\otimes \cdots \otimes Z_{k_n-i_n})(g_{j_1}\otimes \cdots \otimes g_{j_n})|\alpha_{k_1 \cdots k_n}\rangle \\[-1ex]
  && \hspace*{17ex} \langle \beta^c_{{i_1\cdots i_n}}|(g_{j_1}^{\dag c}\otimes \cdots \otimes g_{j_n}^{\dag c})(Z_{i_1-k_1}\otimes \cdots \otimes Z_{i_n-k_n})(g_{j_1}^c\otimes \cdots \otimes g_{j_n}^c)|\beta^c_{k_1 \cdots k_n}\rangle\\[1ex]
 &=& \sum_{i_1\cdots i_n,k_1\cdots k_n} \langle \alpha_{i_1\cdots i_n}|\langle\beta^c_{{i_1\cdots i_n}}| \mathbb{E}_{j_1}[(g_{j_1}^{\dag}Z_{k_1-i_1}g_{j_1})\otimes(g_{j_1}^{\dag c}Z_{k_1-i_1}^{c}g_{j_1}^c)] \otimes \cdots \\[-1ex]
 &&  \hspace*{27ex}\cdots \otimes \mathbb{E}_{j_n}[(g_{j_n}^{\dag}Z_{k_n-i_n}g_{j_n})\otimes(g_{j_n}^{\dag c}Z_{k_n-i_n}^{c}g_{j_n}^c)] \, |\alpha_{k_1 \cdots k_n}\rangle|\beta^c_{k_1 \cdots k_n}\rangle \\
 &=& \<x|A^{\otimes n}|x\> \,.
\end{eqnarray*}

As in the one-shot case, it suffices to show that $\langle x|A^{\otimes n}|x\rangle=0$ implies $\ket{x}=0$.

Note that $A \geq I\otimes(I-\Phi)$, so, $A^{\otimes n} \geq I^{\otimes n} \otimes(I-\Phi)^{\otimes n}$.
Therefore, $$0 = \<x|A^{\otimes n}|x\> \geq \<x|I^{\otimes n} \otimes (I-\Phi)^{\otimes n}|x\> \geq 0$$ so,
$\<x|I^{\otimes n} \otimes (I-\Phi)^{\otimes n}|x\> = 0$.
Equivalently, $\forall i_1\cdots i_n$,
$\Tr[{\alpha_{i_1\cdots i_n}}\otimes{\beta_{i_1\cdots i_n}^c}(I-\Phi)^{\otimes n}]=0$.
Finally, ${\alpha_{i_1\cdots i_n}}\otimes{\beta_{i_1\cdots i_n}^c}$ is a matrix with
positive partial transpose.
According to the following lemma \ref{lemupb}, ${\alpha_{i_1\cdots i_n}}\otimes{\beta_{i_1\cdots i_n}^c}=0$.
So, at most one of $|\alpha_{i_1\cdots i_n}\>$ and $|\beta_{i_1\cdots i_n}^c\>$ can be nonzero
and $\ket{x}=0$.

This completes the proof of Theorem \ref{th1}. \hfill $\blacksquare$

\begin{lemma}\upshape \cite{YuDuanYing14}
\label{lemupb}
For all positive integer $n$, there is no non-zero matrix $M$ satisfying $M \geq 0$, $M^{\Gamma}\geq 0$,
and $\Tr(M(I-\Phi)^{\otimes n})=0$.
\end{lemma}
This lemma was proved in \cite{YuDuanYing14}. We include a proof here to be self-contained.

{\bf Proof.} Suppose by contradiction, there is such a matrix $M$ satisfying those conditions.
Let
\be
\label{twirl}
N=\int\limits_{U} U M U^{\dag}dU,
\ee
where $U$ ranges over all unitaries of the form
$\otimes_{k=1}^n (U_k\otimes U_k^c)$,
and $U_k$ ranges over all unitaries for each $k$, and each $U_k\otimes U_k^c$
acts on the system corresponding to the $k^{\rm th}$ copy of $I-\Phi$.
Note that $N$ satisfies the same properties as $M$, because
the operation in Eq.~(\ref{twirl}) is completely positive (so $N\geq0$), trace preserving (so $N \neq 0$),
PPT preserving (so $N^\Gamma \geq 0$), and finally $\Tr(N(I-\Phi)^{\otimes n})=0$.
Additionally, $N$ satisfies the property that
there are non-negative $p_k$ such that
\[
N=\sum_{R_k\in \mathcal{R}} p_k R_k,
\]
where $\mathcal{R}=\{\Phi,I-\Phi\}^{\otimes n}\setminus \{(I-\Phi)^{\otimes n}\}$.

Since $\Phi^{\Gamma}$ has both strictly positive and strictly negative eigenvalues,
there exists a nonzero $Q$ such that $Q \geq 0$ and $\Tr(Q\Phi^{\Gamma})=0$. Thus,
$$
r:=\Tr(Q(I-\Phi)^{\Gamma})=\Tr(Q) - \Tr(Q\Phi^{\Gamma})=\Tr(Q)>0.
$$

Then the following holds
\begin{equation}\label{tensor}
\Tr_{1,2\cdots,n-1}[(Q^{\otimes n-1}\otimes I)(\sum_{R_k\in \mathcal{R}} p_k R_k^{\Gamma})]\geq 0,
\end{equation}
where $\Tr_{1,2\cdots,n-1}$ denotes the partial trace operation on the first $n-1$ parties
(because the above is a completely positive map to $N^{\Gamma}=\sum_{R_k\in \mathcal{R}} p_k R_k^{\Gamma}\geq 0$.)
Eq. (\ref{tensor}) implies that for $R_k=(I-\Phi)^{\otimes n-1} \otimes \Phi$, we have $r^{n-1}p_k \Phi^\Gamma\geq 0$
which implies that $p_k=0$.  Permuting the systems gives $p_l=0$ for any $R_l$ with $n-1$ tensor factors of
$(I-\Phi)$.  Finally, we can recursively prove that $p_k=0$ for any $R_k\in\mathcal{R}$ with $n-2$ tensor factors,
$n-3$ tensor factors etc.  So, $N=0$ which is a contradiction.  \hfill $\blacksquare$

\section{Conclusion}
In this paper, we show an extreme separation between zero-error
quantum capacity and the private capacity by demonstrating for a class
of channels that the private capacity is maximum given the output
dimension, while there is no ability to transmit even one-qubit with
any finite number of channel uses, when no error can be tolerated. We
hope techniques from our work can be used to study the zero-error
capacity of other channels.

\section{Acknowledgements}

\vspace*{-2ex}

This work is supported by NSERC, NSERC DAS, CRC, and CIFAR.

\vspace*{-1ex}

\newcommand{\etalchar}[1]{$^{#1}$}

\section{Appendix}

For a channel $\mathcal{E}$ with Kraus representation
$\mathcal{E}(\rho)=\sum E_i \rho E_i^{\dag}$ (see \cite{NC00} for
example), its noncommutative graph \cite{Duan09,DuanSeveriniWinter13}
$G(\mathcal{E})$ is defined as the subspace spanned by
$\{E_i^{\dag}E_j:i,j\}$.  In \cite{Shirokov-Shulman-15},
superactivation on the zero-error quantum capacity is studied.
For a channel $\mathcal{E}$ with no one-shot zero-error quantum capacity, 
if one of the following conditions hold, $\mathcal{E}$
cannot be superactivated with any other channel.  

a) $G(\mathcal{E})$ contains a maximal commutative *-subalgebra whose
dimension is the dimension of the input state space. In other words, 
$G(\mathcal{E})$ contains all matrices which are diagonal with respect
to some basis.

b) $G(\mathcal{E})$ is an algebra. In other words, it is closed under matrix production.

In particular, an inductive argument than implies that $\mathcal{E}$ 
has no zero-error quantum capacity.

In this Appendix, we will show that $G(\mathcal{N}_d)$ violates both
conditions, so, our result cannot be inferred from
\cite{Shirokov-Shulman-15}.

First, we observe that the Kraus space of $\mathcal{N}_d$ is spanned by 
$$\{(I\otimes \langle k|)P(I\otimes V) \otimes \ket{V}_{V_B}\otimes
\ket{V}_{V_E}: V\in \mathcal{G}, 1\leq k\leq d\}=\{Z_k\otimes \langle
k|V\otimes \ket{V}_{V_B}\otimes \ket{V}_{V_E}:V\in \mathcal{G}, 1\leq k\leq
d \},$$ where $\mathcal{G}$ is a two-design as described in the main text.
Therefore, 
$$
G(\mathcal{N}_d) = {\rm span}
\{Z_{k-l}\otimes V^{\dag}\op{l}{k}V: V\in \mathcal{G}, 1\leq k,l\leq d \}.
$$
For any $k,l$, we can calculate the span of $\{V^{\dag}\op{l}{k}V:
V\in \mathcal{G}\}$ by considering the isomorphism from $M$ to $(I
\otimes M) \Phi (I \otimes M^\dagger)$ where $\Phi = |\Phi\>\<\Phi|$
and $\ket{\Phi}=\frac{1}{\sqrt{d}}\sum_{i}\ket{i}\ket{i}$ and by averaging 
over $V$: 
\begin{eqnarray*}
&&\mathbb{E}_{V\in \mathcal{G}} (V^{\dag}\op{l}{k}V\otimes I) \; \Phi \; (V^{\dag}\op{l}{k}V\otimes I)^{\dag}\\
&=& \mathbb{E}_{V\in \mathcal{G}}(V^{\dag}\otimes V^T)(\op{k}{k}\otimes\op{l}{l})(V^{\dag}\otimes V^T)^{\dag}\\
&=& \frac{1-\delta_{k,l}/d}{d^2-1}(I-\Phi)+{\delta_{k,l} \over d} \; \Phi \,. 
\end{eqnarray*}
Inverting the isomorphism, the span of $\{V^{\dag}\op{l}{k}V: V\in
\mathcal{G}\}$ is the whole matrix space when $k=l$, and is the space of
all traceless matrices when $k\neq l$.
Note that in particular, $G(\mathcal{N}_d)$ does not contain $Z\otimes I$. 

Now, if condition (a) holds, then there are $d^2$ linearly independent
commuting matrices in $G(\mathcal{N}_d)$, which together with
$Z\otimes I$, gives $d^2+1$ linearly independent commuting matrices 
in $\mathcal{B}(\mathbb{C^{d^2}})$, a contradiction.  
To see that condition (b) does not hold, note that $I\otimes Z,
Z\otimes Z^{\dag} \in G(\mathcal{N}_d)$ but their product $Z\otimes
I \notin G(\mathcal{N}_d)$.

\end{document}